\documentclass[superscriptadress,twoside,onecolumn]{revtex4}
\usepackage{eurosym}
\usepackage{bm}
\usepackage[utf8]{inputenc}
\usepackage{amssymb,theorem,xspace}
\usepackage{float}
\usepackage{amsmath}
\usepackage{mathrsfs}
\usepackage{graphicx}
\usepackage{color}
\usepackage{caption}
\usepackage{subcaption}
\begin{document}

\title{On the Fractional Quark-Antiquark Confinement and Symplectic Quantum Mechanics}
\author{M. Abu-Shady}
\email{dr.abushady@gmail.com}
\affiliation{Department of Mathematics and Computer Science, Faculty of Science, Menoufia
University, Shibin Al Kawm, Egypt}

\author{R.R. Luz}
\email{renatoluzrgs@gmail.com}
\affiliation{International Center of Physics, Instituto de Física, Universidade de
Brasília, 70.910-900, Brasília, DF, Brazil }

\author{G.X.A. Petronilo}
\email{gustavopetronilo@gmail.com}
\affiliation{International Center of Physics, Instituto de Física, Universidade de
Brasília, 70.910-900, Brasília, DF, Brazil }

\author{A.E. Santana}
\email{a.berti.santana@gmail.com}
\affiliation{International Center of Physics, Instituto de Física, Universidade de
Brasília, 70.910-900, Brasília, DF, Brazil }

\author{R.G.G. Amorim}
\email{ronniamorim@gmail.com}
\affiliation{International Center of Physics and Gama faculty, Universidade de
Brasília, 72.444-240, Brasília, DF, Brazil }
\affiliation{Canadian Quantum Research Center,\\
204-3002 32 Ave Vernon, BC V1T 2L7  Canada}

\begin{abstract}
Using the formalism of generalized fractional derivatives, a two-dimensional non-relativistic meson system is studied. The
mesons are interacting by a Cornell potential. The system is formulated in the domain of the symplectic quantum mechanics by means of the generalized fractional
Nikiforov-Uvarov method. The corresponding Wigner function and the energy eigenvalues are then derived. The effect of fractional parameters $\alpha$ and $\beta$
with the ground state solution is analyzed through the Wigner function for the charm-anticharm, bottom-antibottom and $b\overline{c}$ mesons.
One of the fundamental achievements of such Cornell model is the determination of heavy quarkonia mass spectra. We have computed these masses and the
present results are in agreement with the experimental data, improving previous theoretical results.
\end{abstract}

\maketitle

\section{Introduction}\label{sec:Introduction}

The advent of the charmonium system  (a bound state of the quark $c$ and its antiparticle $\overline{c}$) and advances in quantum chromodynamics (QCD) have led to a better understanding of hadron physics~\cite{Godfrey}. The resemblance to positronium was evident, thus the term ``quarkonium'' was conceived~\cite{Grosse,Mirjalili}. Since then, various theoretical approaches have been used to consider the phenomenon of quark confinement in QCD, including lattice QCD, heavy quark effective field theory, and potential-dependent models~\cite{Mansour,mann,gasiorowicz,ABUSHADY,griffiths,mabu,%
Maireche,aitchison,Kumar,thomson}.

Non-relativistic potentials are effective in describing heavy-quark systems, capturing the dynamics assumed by QCD. Numerous models exist to describe the spectra of charm-anticharm ($c\overline{c}$) and bottom-antibottom ($b\overline{b}$) states. In this case, the Cornell potential is one of the earliest and most widely studied models in particle physics~\cite{Purohit,chung,soni,belich,petronilo,mabu2018,vega,Mutuk2019,%
Tajik2019,Ferkous2013,Inyang2021,Omugbe2020,Khoka2016,Inyang2020,mabu2,%
Ahmadov2019,Sameer2013,Faizuddin2021,Nahool,Gupta2012,AS_3,As_4}. This potential, which is expressed as
\begin{equation}
V(q) = aq - \frac{b}{q},\label{CornI}
\end{equation}
where $a,b\geq 0$, are constants and $q$ stands for a coordinate separation of quarks in the configuration space. This expression encompasses a one-gluon exchange term akin to one-photon exchange and a confinement (the linear) term resembling to a string. It has been successful in modeling the binding of heavy quarks and, due to its characteristics and applications, has garnered considerable attention. The Cornell potential is particularly relevant for investigating the confined and deconfined phases within hadrons~\cite{RL_1}. Since this potential is a function of points in an Euclidian space,   the Schrödinger equation has been used as a model-dependent quarkonium equation, yielding valuable insights.

In theoretical studies, several techniques have been developed to solve the Schr\"odinger equation for quarkonium systems. For instance, the non-relativistic quark model and the Nikiforov-Uvarov approach was used to investigate the  masses and thermodynamic properties of heavy mesons~\cite{mabu,As_4}. By considering  the generalized Bopp-shift method and standard perturbation theory~\cite{Maireche}, a modified Cornell potential was introduced to solve analytically the three-dimensional Schrödinger equation for heavy quarkonium systems. That work provided important informations about the modified mass and other characteristics of quarkonium systems influenced by the modified Cornell potential at finite temperature. Noncommutative space-time effects on the Cornell potential in heavy quarkonium systems were also explored~\cite{Mirjalili}. The radial Schrödinger equation with an exponential in the Cornell potential was analytically solved by using a series  expansion method. In this case, the corresponding bound-state energy spectra was obtained for specific systems~\cite{Etebong}. Additionally, the bound-state energy spectra of heavy systems was derived by combining the Kratzer and screened Coulomb potentials and by solving the radial Schrödinger equation~\cite{Ibekwe}. The extension of the Cornell potential was achieved by introducing a quadratic correction (a harmonic oscillator potential). The resulting Schrödinger equation in $N$ dimensions was solved for meson systems involving $b\overline{c}$ and $c\overline{c}$ quark pairs, yielding the mass spectra~\cite{mabu1}. Finally,  the linear term in the Cornell potential plus a modified Yukawa potential was also considered~\cite{Purohit}. This led to   bound-state solutions approaching  to those ones derived from the Klein-Gordon equation in three-dimensional space. Through the Nikiforov-Uvarov formulation, the Klein-Gordon equation was analytically solved, providing the energy eigenvalues and wave functions for heavy meson systems.

However, studies on the heavy quark-antiquark interaction within the background of fractional derivatives in the phase space representation remain a little explored in the literature. This is the case despite the fact that  fractional calculus include successful applications in damping oscillator  systems, quantization of dissipative systems, sketch of heavy mesons energy spectra and elements of the intricate standard model~\cite{Jamel,Oldham,Podlubny,Baleanu}. Considering those aspects,  a generalized fractional derivative  was  developed~\cite{AS_1}. These definitions gave rise to satisfactory results, when applied to  interacting models~\cite{mabu2,abu-shad2022,abu-shad,rluz}.

Along this perspective, the main purpose of this work is to study the charm-anticharm, bottom-antibottom and bottom-anticharm mesons through a generalized fractional derivative by using the framework of the symplectic quantum mechanics and the extended fractional Nikiforov-Uvarov method. We analyse the behavior of the Wigner function, what is defined in phase space, for the ground state of the heavy mesons system by observing the effects of fractional parameters. In addition,  the heavy quarkonia ($c\overline{c}$, $b\overline{b}$ and $b\overline{c}$ mesons) spectra is determined. The motivation toward using Wigner function is that, in recent times, it has been studied for the strongly bound system in numerous  models~\cite{jana}. The Wigner function is an excellent tool to investigate the nature of a  quantum-system state, such as chaoticity and non-classicality, which are important concepts  in quantum computing and quantum information, but also  for  QCD, in particular for studying the quark distribution~\cite{jana,lorce,ojha,Ferry}.

The structure of this work is the following. In the Section~\ref{sqm}, some aspects of the Schrödinger equation represented in phase space and  the generalized fractional Nikiforov-Uvarov method are shortly reviewed  to fix the notation.
Section~\ref{FSE} introduces the fractional quark-antiquark structure for the Symplectic Schrödinger equation.
 Section~\ref{dr} is devoted to the discussion of outcomes. In Section~\ref{fr}, final concluding remarks  are presented.

\section{Symplectic Quantum Mechanics and Nikiforov-Uvarov Method: an Overview}\label{sqm}
In this section, a brief review of the symplectic quantum mechanics and the generalized fractional Nikiforov-Uvarov method are presented. In the next sections both structures are explored to study properties of mesons.
\subsection{Non-relativistic Symplectic Quantum Mechanics}
Consider  $\Gamma$ a set of points denoted by $(q,p)$, defined in the cotangent space $T^*\mathbb E$,  where $q$ is a coordinate in the configuration space defined in the Euclidian space, $\mathbb E$.  When equipped with the symplectic 2-form $w=dq\wedge dp$, then $T^*\mathbb E$ is called  phase space. A Hilbert space is
introduced in $\Gamma$ by taking the set of $C^\infty $-funtions $\phi(q,p)$ fulfilling the property
$
\int  dqdp\phi ^{*}(q,p)\phi (q,p)< \infty.
$ 
This Hilbert space is denoted by $\mathcal{H}(\Gamma)$.
A base in $\mathcal{H}(\Gamma)$ is given by  $| q,p\rangle$, such that the dual is $\langle q,p |$.  The completeness property is
$\int  dqdp| q,p\rangle \langle q,p |=1$. A general vector in $\mathcal{H}(\Gamma)$ is written as  $| \phi\rangle$, with the dual $\langle \phi |$, such that $\phi(q,p)= \langle q,p | \phi\rangle$.

The Hilbert space $\mathcal{H}(\Gamma)$ is a carrier space for Lie symmetries.  Indeed, unitary mappings,   $U(\alpha )$ in $\mathcal{H}(\Gamma )$, are defined as $U(\alpha )=\exp
(-i\alpha \widehat{A})$, where
\begin{align*}
\widehat{A}& =A(q,p)\star =A(q,p)\exp \left[ \frac{i\hbar }{2}\left(\frac{%
\overleftarrow{\partial }}{\partial q}\frac{\overrightarrow{\partial }}{%
\partial p}-\frac{\overleftarrow{\partial }}{\partial p}\frac{%
\overrightarrow{\partial }}{\partial q}\right)\right] \\
& =A\left(q+\frac{i\hbar }{2}\partial _{p},p-\frac{i\hbar }{2}\partial _{p}\right)
\end{align*}
and the star (Weyl) product, $\star$, is given by
$
\star = \exp \left[ \frac{i\hbar }{2}\left(\frac{%
\overleftarrow{\partial }}{\partial q}\frac{\overrightarrow{\partial }}{%
\partial p}-\frac{\overleftarrow{\partial }}{\partial p}\frac{%
\overrightarrow{\partial }}{\partial q}\right)\right].
$

The basic functions in $\Gamma, $  $q$ and
$p$ (3-dimensional Euclidean vectors),  lead to the hat operators
\begin{eqnarray}
\widehat{P} &=&p\star =p-\frac{i}{2}\partial_{q},  \label{g1} \\
\widehat{Q} &=&q\star =q+\frac{i}{2}\partial_{p}.  \label{g2}
\end{eqnarray}%
Here, we are using the Planck constant  taken as $\hbar = 1$. A symplectic framework for quantum mechanics is established by identifying, first, the Heisenberg commutation relation, which is given by $\left[ \widehat{Q},\widehat{P}\right] =i$. The   following operators are introducied:
$\widehat{K}_{i} = m\widehat{Q}_{i}-t\widehat{P}_{i}$,
$\widehat{L}_{i} = \epsilon _{ijk}\widehat{Q}_{j}\widehat{P}_{k}$,
$\widehat{H} = \frac{\widehat{P}^{2}}{2m}$,
with ``$m$'' representing a central extension standing for the mass . The operators  $\widehat{P},\widehat{K},\widehat{L}$ and $\widehat{H}$ are the generators of Galilei symmetries, standing  for translation, Galilean boosts, rotation, and time translation, respectively.

Using the time-translation generator, $\widehat{H}$, the time displacement of a wave function
in phase space is given by
$
\psi(q,p,t)=e^{\widehat{H}(t-t_0}\psi(q,p,t_0).
$
This lead to
\begin{equation}
\partial_t \psi(q,p;t)=\widehat{H}(q,p) \psi(q,p;t), \label{SSEqI}
\end{equation}
which is the so-called symplectic Schrodinger equation~\cite{oliveira2004}.

The physical interpretation of this formalism is obtained by establishing the
connection of $\psi(q,p,t)$ with a function a Wigner function, a quasi distribution of probability, $f_W$, that is~\cite{oliveira2004,paiva2018,dessano,paiva2020,Martins}
$
f_W(q,p,t)=\psi(q,p,t)\star\psi^\dagger(q,p,t).
$
Due to this result, the wave function $\psi(q,p,t)$ is interpreted as a quasi amplitude of probability~\cite{RL_1,paiva2020,campos2017,campos2018}. In the next section, the extended Nikiforov-Uvarov method is  addressed in the context of the fractional derivatives.

\subsection{The Generalized Fractional Nikiforov-Uvarov Method} \label{NU}

The generalized Nikiforov-Uvarov method represents an expansion of the standard Nikiforov-Uvarov method, with both techniques primarily
employed within the realm of quantum mechanics. This approach lies in the determination of eigenvalues and eigenfunctions for a range
of equations, including Schrödinger and Dirac equations, as well as equations susceptive to transformation into hypergeometric form, that is~\cite{Karayer,Nikiforov,Jamel,soleiman,MAbu2016,AS_4,Shady,kaabar},
\begin{equation}
    D^{\alpha}\left [ D^{\alpha}\psi(s) \right ]+\frac{\widetilde{\tau }(s)}{\sigma (s)}D^{\alpha}\psi(s)+\frac{\widetilde{\sigma }(s)}{\sigma^{2}(s)}\psi(s)=0. \label{gf0}
\end{equation}
Here, $\sigma (s)$ and $\widetilde{\sigma }(s)$ are polynomials of maximum second degree of $\alpha$ and $2\alpha$, respectively, and $\widetilde{\tau}(s)$ has a maximum degree of $\alpha$, such that
\begin{align}
    D^{\alpha}\psi(s)&=Is^{1-\alpha}\psi'(s), \label{gf1}\\
D^{\alpha}\left [ D^{\alpha}\psi(s) \right ]&=I^{2}\left [ (1-\alpha)s^{1-2\alpha}\psi'(s)+s^{2-2\alpha}\psi''(s) \right ], \label{gf2}
\end{align}
where $I=\frac{\Gamma(\beta)}{\Gamma(\beta-\alpha+1)}$ and the fractionaries parameters satisfies the condition, $0< \alpha \leq 1$, $0< \beta \leq 1$. By taking Eqs.~\eqref{gf1}~and~\eqref{gf2} into Eq.~\eqref{gf0}, it follows that
\begin{equation}
    \psi''(s)+ \frac{\widetilde{\tau}_f(s)}{\sigma_f(s)}\psi'(s)+\frac{\widetilde{\sigma}_f(s)}{\widetilde{\sigma}_{f}^{2}(s)}\psi(s)=0, \label{gf3}
\end{equation}
where the subscript $f$ denotes fractional and
\begin{align}
    \widetilde{\tau}_{f}(s)&=(1-\alpha)s^{-\alpha}\sigma(s)+I^{-2}\widetilde{\tau}(s), \\
    \sigma_{f}(s)&=s^{1-\alpha}\sigma(s), \\
\widetilde{\sigma}_{f}(s)&=I^{-2}\widetilde{\sigma}(s).
\end{align}

 Writing
\begin{equation}
    \psi(s) = \phi(s)y(s),
\end{equation}
then Eq.~\eqref{gf3} leads to the following the hypergeometric equation,
\begin{align}
    \sigma_{f}(s)y''(s)+\tau_{f}(s)y'(s)+\lambda y(s)=0,
\end{align}
where $\phi(s)$ is represented as the logarithm derivative, i.e.,
\begin{align}
\frac{\phi'(s)}{\phi(s)} =\frac{\pi_{f}(s)}{\sigma_{f}(s)}, \label{pearsonEq}\\
\end{align}
and
\begin{equation}
    \tau_{f}(s) =\widetilde{\tau}_{f}(s)+2\pi_{f}(s),
\end{equation}
For bound solutions, it is required that
\begin{equation}
    \tau'(s) < 0.
\end{equation}
The equation of eigenvalues is given by
\begin{equation}
    \lambda = \lambda_{n} = -n\tau'_{f}(s)-\frac{n(n-1)}{2}\sigma''_{f}(s), \quad (n=0,1,2,...)
\end{equation}
and $y(s)$ is a hypergeometric type function, whose polynomial solutions are obtained from Rodrigues' relation
\begin{equation}
    y(s)=y_{n}(s)=\frac{B_n}{\rho (s)}\frac{d^{n} }{d s^{n}}\left [ \sigma_{f}^{n}(s)\rho(s) \right ],
\end{equation}
where $B_{n}$ is a normalization constant and $\rho(s)$ is a weight function which satisfies the following equation
\begin{align}
    \frac{d\omega(s)}{ds} = \frac{\tau(s)}{\sigma_{f}(s)}\omega(s), \label{pearsonEquation1}
\end{align}
such that $\omega(s)=\sigma_{f}(s)\rho(s)$. The function $\pi_{f}(s)$ is defined as
\begin{equation}
    \pi_{f}(s)=\frac{\sigma'_{f}(s)-\widetilde{\tau}_{f}(s) }{2}\pm \sqrt{\left ( \frac{\sigma'_{f}(s)-\widetilde{\tau }_{f}(s)}{2} \right )^{2}-\widetilde{\sigma}_{f}(s)+K\sigma_{f}(s)}, \label{gf4}
\end{equation}
and
\begin{equation}
    \lambda=K+\pi'_{f}(s), \label{gf5}
\end{equation}
where  $\pi_{f}(s)$ is a first-degree polynomial. If the expressions under the square root are squares of expressions, the values of $K$ in Eq.~\eqref{gf4} may be determined. This is possible if its discriminate is zero~\cite{mabu,Shady,MAbu2016,ABUSHADY}. In the following, this methodology is implemented  to solve symplectic Schrödinger equation into domain of the generalized fractional derivatives for the Cornell potential.

\section{Fractional Quark-antiquark System and Symplectic Schrödinger Equation}\label{FSE}

Consider a bound state of the heavy quark-antiquark such as $c\overline{c}$, $b\overline{b}$ mesons of mass $m$. These two particles interact
with one another by means of phenomenological Cornell potential, given in Eq.~(\ref{CornI}). The Taylor series  around a point $q_0$ can be used to derive the Cornell potential, which is written, up to second order approximation, as
\begin{equation}
V(q)=-\frac{3b}{q_0}+\bigg(a+\frac{3b}{q_0^2}\bigg)q-\frac{b}{q_0^3}q^2,
\end{equation}
where the $a,b$ are purely phenomenological constants of the model, and $q_0$ is one specific relative spatial coordinate between the two quarks. The steady symplectic Schrödinger equation is obtained form Eq.~(\ref{SSEqI}), leading explicitly to
\begin{equation}
    H\star\psi=\frac{p^2}{2m}\star\psi +\bigg(a+\frac{3b}{q_o^3}\bigg)q\star\psi-\frac{b}{q_0^3}
    q^2\star\psi=\Big(E+\frac{3b}{q_0}\Big)\psi.\label{eq_cp}
\end{equation}
Here, $E$ is the eigenvalue of $\widehat{H}=H\star$ and  $m=\frac{m_{q}m_{\overline{q}}}{m_q + m_{\overline{q}}}$ stands for the reduced mass for the quarkonium particle ($c\overline{c},b\overline{b}$ and $b\overline{c}$). This  Eq.~\eqref{eq_cp} is rewritten as
\begin{equation}
    H\star\psi=\frac{p^2}{2m}\star\psi +\lambda q\star\psi-\sigma q^2\star\psi=E^\prime\psi,\label{eq_cp1}
\end{equation}
where $\lambda=\bigg(a+\dfrac{3b}{q_o^3}\bigg)$, $\sigma=\dfrac{b}{q_o^3}$ and $E^\prime=E+\dfrac{3b}{q_o}$.

Using Eq.~(\ref{g1}), and Eq.~(\ref{g2}) into Eq.~\eqref{eq_cp1}, the steady-state symplectic Schrödinger equation reads
\begin{equation}
\frac{1}{2m}\left( p^{2}-ip\partial _{q}-\frac{1}{4}\partial _{q}^{2}\right)
\psi +\lambda \left( q+\frac{i}{2}\partial _{p}\right) \psi-\sigma\left( q^{2}-iq\partial _{p}-\frac{1}{4}\partial _{p}^{2}\right) =E^\prime\psi .
\label{D_3}
\end{equation}%
(we are using  $\hbar =1$). By using the transformation
$z =\frac{p^{2}}{2m}+\lambda q-\sigma q^2$, one obtain~\cite{RL_1}
\begin{equation}
\bigg(z-\frac{\lambda^2}{4\sigma}\bigg)\frac{\partial^2\psi}{\partial z^2}+\frac{\partial\psi}{\partial z}+\frac{2m}{\sigma}\bigg(z-E^\prime\bigg)\psi=0
\end{equation}%
or
\begin{equation}
 \frac{\partial^2\psi}{\partial\omega^2}+   \frac{1}{   \omega}\frac{\partial\psi}{\partial\omega}+\frac{2m}
 {\omega\sigma}\bigg(\omega+\frac{\lambda^2}
 {4\sigma}-E^\prime\bigg)\psi=0,\label{eq-se1}
\end{equation}
where $\omega=z-\frac{\lambda^2}{4\sigma}$.

Considering the ansatz
\begin{equation}
 \psi=\omega^{-\frac{1}{2}}R(\omega)\label{eq-r1},
\end{equation}
the expressions for the derivatives are given by
\begin{eqnarray}
\frac{\partial\psi}{\partial\omega}=\omega^{-\frac{1}{2}}\frac{\partial R}{\partial \omega}-\frac{1}{2}\omega^{-\frac{3}{2}}R(\omega),\label{eq-r2}\\
\frac{\partial^2\psi}{\partial \omega^2}=\omega^{-\frac{1}{2}}\frac{\partial^2 R}{\partial \omega^2}-\omega^{-\frac{3}{2}}\frac{\partial R}{\partial \omega}-\frac{3}{4}\omega^{-\frac{5}{3}}R(\omega).\label{eq-r3}
\end{eqnarray}
By substituting Eqs.~\eqref{eq-r1},~\eqref{eq-r2}~and~\eqref{eq-r3} into Eq.~\eqref{eq-se1}, after some calculations, one obtains
\begin{equation}
  \frac{\partial^2 R}{\partial\omega^2}+\frac{1}{4\sigma\omega^2}\bigg(\sigma+8m\omega\Big(\omega+\frac{\lambda^2}{4\sigma}-E^\prime\Big)\bigg)R(\omega)=0.\label{eq-r4}
\end{equation}

Using the generalized fractional derivative framework~\cite{AS_1},  Eq.~\eqref{eq-r4} leads to
\begin{eqnarray}
D^\alpha D^\alpha R(\omega)+\frac{1}{4\sigma\omega^2}\bigg(\sigma+8m\omega\Big(\omega+\frac{\lambda^2}{4\sigma}-E^\prime\Big)\bigg)R(\omega)=0.\label{eq-g0}
\end{eqnarray}
From Eq.~\eqref{gf2}, we have
\begin{equation}
    D^\alpha D^\alpha R(\omega)=\bigg(\frac{\Gamma(\beta)}{\Gamma(\beta-\alpha+1)}\bigg)^2\Bigg[\omega^{2-2\alpha}\frac{\partial^2 R}{\partial \omega^2}+(1-\alpha)\omega^{1-2\alpha}\frac{\partial R}{\partial\omega}\bigg].\label{eq-g1}
\end{equation}
Considering this equation, Eq.~\eqref{eq-g0} reads
\begin{equation}
    \frac{d^{2}R}{d\omega^2}+\frac{(1-\alpha)}{\omega^{'}}\frac{dR}{d\omega }+\frac{1}{4\sigma A^2 \omega^{'2}}\left [ \sigma + 8m\omega^{'2\alpha}+\frac{\lambda^{'2}8m \omega^{'\alpha}}{4\sigma }-8m\omega^{'\alpha}E \right ]R(\omega )=0,  \label{eq-g2}
\end{equation}
where
\begin{equation}\label{eq-A}
    A = \frac{\Gamma (\beta)}{\Gamma (\beta - \alpha + 1)} \quad ; 0< \alpha \leq 1 \quad ; 0< \beta \leq 1.
\end{equation}
By comparing Eq.~\eqref{gf3} and Eq.~\eqref{eq-g2}, we define the following  equations in order to use the Nikiforov-Uvarov (NU) method~\cite{Karayer}:
\begin{equation}
    \tilde{\tau_f}(\omega)=1- \alpha \quad ; \quad \sigma_{f}=\omega \label{eq-g3}
\end{equation}
and
\begin{equation}
    \tilde{\sigma_f}=\frac{1}{4\sigma A^2}\left [ \sigma + 8m^{'}\omega^{'2\alpha}+\frac{8m\omega^{'\alpha}\lambda^{'2} }{4\sigma }-8m\omega^{'\alpha}E \right ]. \label{eq-g4}
\end{equation}
The  Nikiforov-Uvarov approach is carried out by considering  Eq.~\eqref{gf4} to write $\pi_{f}$ as
\begin{equation}
    \pi_{f}=\frac{\sigma^{'}_{f}(\omega^{'})-\pi_{f}(\omega^{'})}{2}\pm \sqrt{\left (\frac{\sigma_{f}(\omega^{'})-\pi_{f}(\omega^{'})}{2}  \right )^{2}-\tilde{\sigma }_{f}(\omega^{'})+K\sigma_{f}(\omega)}. \label{eq-g5}
\end{equation}
By substituting   Eqs.~\eqref{eq-g3}~and~\eqref{eq-g4} into Eq.~\eqref{eq-g5},   $\pi_{f}$ reads
\begin{equation}
    \pi_{f}=\frac{\alpha}{2}\pm \sqrt{\frac{\alpha^2}{4}-A_{1}-A_{2}\omega^{'\alpha}-A_{3}\omega^{'2\alpha} +\omega K }.
\end{equation}
The constant parameter $K$ is be determined by utilizing the condition that the expression under the square root has a double zero, i.e., its discriminant is equal to zero. Hence, it follows that $K=C_{1}\omega^{\alpha - 1}$. As a consequence,  $\pi_{f}$ is given by
\begin{equation}
    \pi_{f}=\frac{\alpha}{2}\pm \sqrt{\frac{\alpha^2}{4}-A_{1}+(C_1-A_2)\omega^{\alpha}-A_3 \omega^{'2\alpha} }, \label{eq-g6}
\end{equation}
where
\begin{equation}
    C_{1}=A_{2}\pm \sqrt{4A_{3}\left ( A_{1}-\frac{\alpha^2}{4} \right )}, \label{eq-g7}
\end{equation}
with
\begin{align}
    A_{1}&=-\frac{1}{4A^2},  \\
A_{2}&=\frac{1}{4\sigma A^2}\left ( \frac{8m\lambda^{'2}}{4\sigma }-8mE\right ),  \\
A_{3}&=\frac{8m^{'}}{4\sigma A^2}.  \label{eq-g12}
\end{align}

From Eqs.~(\ref{eq-g6})~and~(\ref{eq-g7}), there are four possible forms of $\pi_{f}$, which are the following:
\begin{align}
    \pi_{f}=\frac{\alpha}{2}\pm \begin{cases}
 \sqrt{A_3}\omega^{'\alpha}+ \sqrt{A_1 - \frac{\alpha^2}{4}}& \text{ for } k_{1}=A_{2}+ \sqrt{4A_3\left ( A_1 - \frac{\alpha^2}{4} \right )}, \\
 \sqrt{A_3}\omega^{\alpha} - \sqrt{A_1 - \frac{\alpha^2}{4}} & \text{ for } k_{1}=A_{2}- \sqrt{4A_3\left ( A_1 - \frac{\alpha^2}{4} \right )}.
\end{cases}
\end{align}
We select $\pi_{f}$ as~\cite{Karayer},
\begin{equation}
    \pi_{f} = \frac{\alpha}{2} - \sqrt{A_{3}}\omega^{\alpha} + \sqrt{A_1 - \frac{\alpha^2}{4}}. \label{eq-g8}
\end{equation}
Defining~\cite{Karayer}
\begin{equation}
    \pi_{f}(\omega)= \frac{1}{2}\left ( \tau_{f}(\omega) - \tilde{\tau_f}(z) \right ), \label{eq-g9}
\end{equation}
and using Eqs.~\eqref{eq-g8}~and~\eqref{eq-g3}, we obtain
\begin{equation}
    \tau_{f}(\omega)= -2\sqrt{A_3}\omega^{\alpha} + 2\sqrt{A_1 - \frac{\alpha^2}{4}} + 1.
\end{equation}
From Eq.~\eqref{gf5}, we define
\begin{align}
    \lambda (\omega)&= K(\omega^{'}) + \pi_{f}^{'}(\omega^{'}), \nonumber \\
    \lambda (\omega)&= \left [ A_2 - \sqrt{4A_3 \left (A_1 - \frac{\alpha^2}{4}  \right )} \right ]\omega^{\alpha - 1}. \label{eq-g10}
\end{align}
and also
\begin{align}
    \lambda_{n}(\omega^{'})&=-\frac{n}{2}\pi_{f}^{'}(\omega^{'}) - \frac{n(n-1)}{\sigma^2}\sigma_{f}(\omega), \nonumber \\
    \lambda_{n}(\omega)&=n\alpha \sqrt{A_3}\omega^{\alpha - 1}.\label{eq-g11}
\end{align}
From  the right hand sides of Eqs.~\eqref{eq-g10}~and~\eqref{eq-g11} we have
\begin{equation}
    A_2 - \sqrt{4A_3 \left ( A_1 - \frac{\alpha^2}{4} \right )}=n \alpha \sqrt{A_3}. \label{eq-g13}
\end{equation}
With Eqs.~\eqref{eq-g12}, Eq.~\eqref{eq-g13} provides the energy eigenvalues in the fractional space, that is,
\begin{equation}\label{eq-e1}
    E^{'}=\frac{\lambda^2}{4\sigma}-\frac{\sigma A^2}{2m}\left [ n\alpha \sqrt{\frac{m}{4\sigma m^2}} + \sqrt{\frac{m}{\sigma A^2}\left ( \frac{1}{A^2}- \alpha^2 \right )}\right ] ,
\end{equation}
where $A = \frac{\Gamma(\beta)}{\Gamma(\beta - \alpha + 1)}$, $0< \alpha \leq 1$, $0< \beta \leq 1$. Since   $E'$ is real-valued in  Eq.~(\ref{eq-e1}),  then the following condition is derived,
\begin{equation}\label{eq-e2}
\frac{1}{A^2}- \alpha^2\geq 0.
\end{equation}
Once $A$ is defined by Eq.~(\ref{eq-A}), the choice of $\alpha$ and $\beta$ is arbitrary; i.s., it has to satisfy  the condition
\begin{equation}\label{eq-e3}
\left(\frac{\Gamma (\beta-\alpha+1)}{\Gamma (\beta)}\right)^2-\alpha^2\geq 0.
\end{equation}

Taking $R(\omega)=\varphi (\omega)y(\omega)$, then from Eq.~\eqref{pearsonEq}, $\phi(\omega)$ is a solution of the equation
\begin{equation}
    \frac{\varphi^{'}(\omega)}{\varphi(\omega)}=
    \frac{\pi_{f}(\omega)}{\sigma_{f}(\omega)}. \label{logEq}
\end{equation}
Substituted $\pi_{f}(\omega)$ and $\sigma_{f}(\omega)$ into equation~\eqref{logEq}, the result is
\begin{equation}
    \varphi = \omega^{\frac{\alpha}{2}+\sqrt{A_1 - \frac{\alpha^2}{4}}}e^{-\frac{\sqrt{A_3}\omega^{\alpha}}{\alpha}}. \label{eq-h}
\end{equation}
Furthermore, the other part of the field given by the quasi-amplitude of probability $y(\omega)$ is the hypergeometric-type function obtained from Eq.~\eqref{pearsonEquation1} as follows
\begin{equation}
    \frac{d}{d\omega}\left ( \sigma_{f}(\omega)\rho(\omega) \right )=\tau(\omega)\rho(\omega).
\end{equation}
So that
\begin{equation}
  \sigma_{f} \frac{d\rho}{d\omega} + \rho(\omega)\frac{d\sigma_f}{d\omega} = \tau(\omega)\rho(\omega).
\end{equation}
By considering $\sigma_{f}(\omega)=\omega$, we have
\begin{equation}
    \omega \frac{d\rho}{d\omega} + \left (-\sqrt{A_3}\omega^{\alpha} + 2\sqrt{A_1 - \frac{\alpha^2}{4}}+1  \right )\rho.
\end{equation}
By solving the above equation, we obtain
\begin{equation}
 \rho = \omega^{2\sqrt{A-\alpha^{2} / 4}}exp\left [-\frac{A_2}{\alpha}\omega^{\alpha}  \right ].    \label{eq-h1}
\end{equation}
Then $y_{n}(\omega)$ is given by
\begin{equation}
    y_{n}(\omega) = \frac{B_n}{\rho(\omega)}\frac{d^n}{d\omega^n}[\sigma_{f}^{n}(\omega)\rho(z)]. \label{eq-h2}
\end{equation}
With Eq.~\eqref{eq-h1}, Eq.~\eqref{eq-h2} yields
\begin{equation}
    y_{n}(\omega)=B_{n}\omega^{-2\sqrt{A_1 - \frac{\alpha^2}{4}}}e^{2\frac{A_3  }{\alpha}\omega^{\alpha}}\frac{d^n}{d\omega^n}\left [\omega^{n+1\sqrt{A_1 - \frac{\alpha^2}{4}}}e^{-2\frac{\sqrt{A_3}}{\alpha}\omega^{\alpha}}  \right ]. \label{eq-h3}
\end{equation}

From Eq.~\eqref{eq-h} and Eq.~\eqref{eq-h3}, we write
\begin{equation}
    R(\omega)=C_{n}\omega^{\frac{\alpha}{2}-\sqrt{A_1 - \frac{\alpha^2}{4}}}e^{\frac{\sqrt{A_3}}{\alpha}\omega^{\alpha}}\frac{d^n}{d\omega^n}\left [\omega^{n + 2\sqrt{A_1 - \frac{\alpha^2}{4}}}e^{-2\frac{\sqrt{A_3}}{\alpha}\omega^{\alpha}}  \right ].
\end{equation}
Using Eq.~\eqref{eq-r1}, we have
\begin{equation}
    \psi_n(\omega)=C_{n}\omega^{\frac{1+\alpha}{2}-\sqrt{A_1 - \frac{\alpha^2}{4}}}\frac{d^n}{d\omega^n}\left [\omega^{n + 2\sqrt{A_1 - \frac{\alpha^2}{4}}}e^{-2\frac{\sqrt{A_3}}{\alpha}\omega^{\alpha}}  \right ]. \label{solfrac2}
\end{equation}
In this way, the $n$-order Wigner function is given by
\begin{equation}\nonumber
f_W^{(n)}(q,p)=\psi_n(\omega)\star\psi_n^{\ast}(\omega).
\end{equation}
Then the calculation of the $n=0$ Wigner function, up to second order in $hbar$ in the star-product, leads to
\begin{eqnarray}
f_W^{(0)}(q,p)&=&N\omega^{(1+\alpha)}e^{-2\frac{\sqrt{A_3}}{\alpha}\omega^{\alpha}}\nonumber\\
&&-N\sqrt{\frac{1}{A^2}+\alpha^2}\omega^{(1+\alpha)}e^{-2\frac{\sqrt{A_3}}{\alpha}}(\lambda-2\sigma q)\frac{p}{m}, \label{fracWigner}
\end{eqnarray}
where $N$ is a normalization constant which is computed by $\int f_W^{(0)}(q,p)dqdp=1$. These results are considered in detail in the next section.

\section{Discussion of Results} \label{dr}
\begin{figure}[H]
\centering
\includegraphics[scale=0.29]{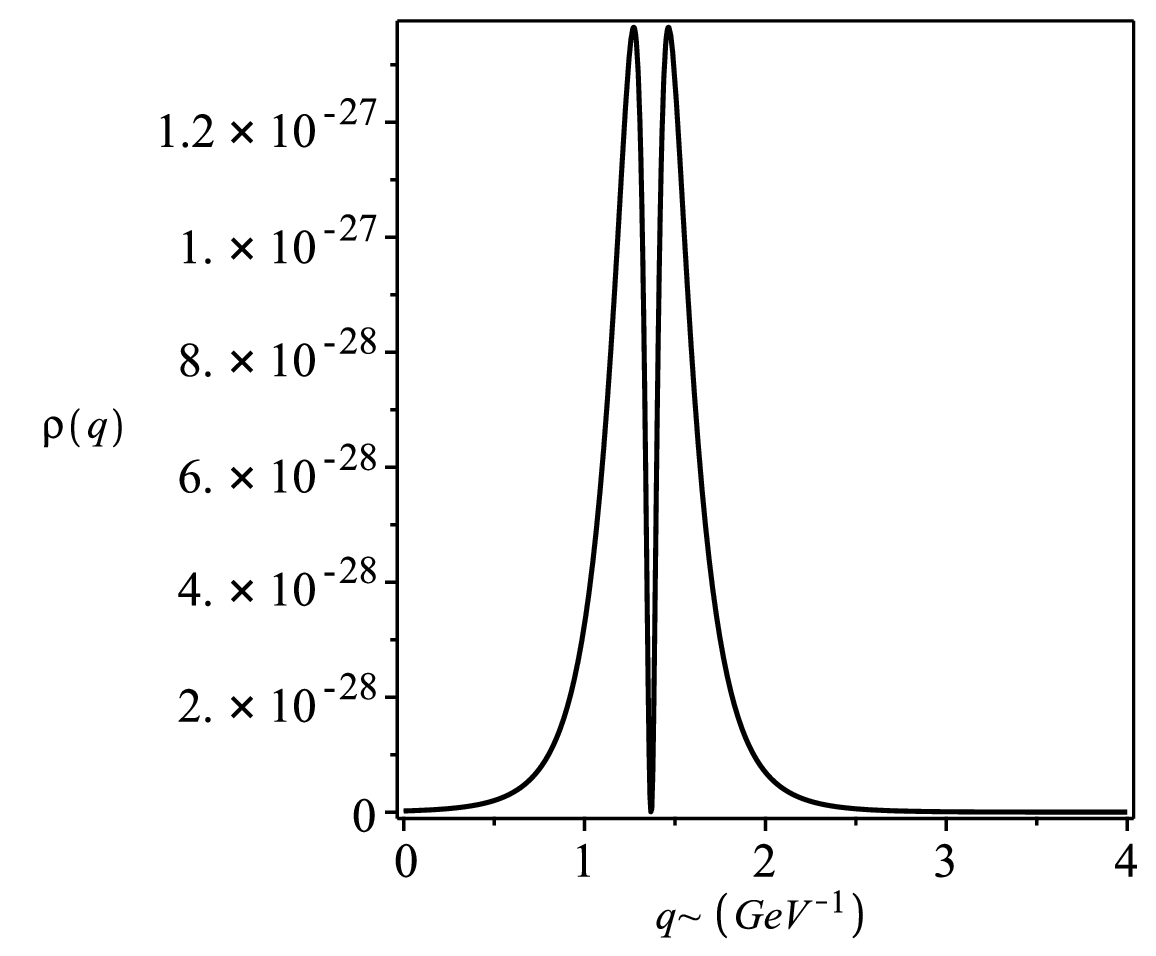}
\includegraphics[scale=0.29]{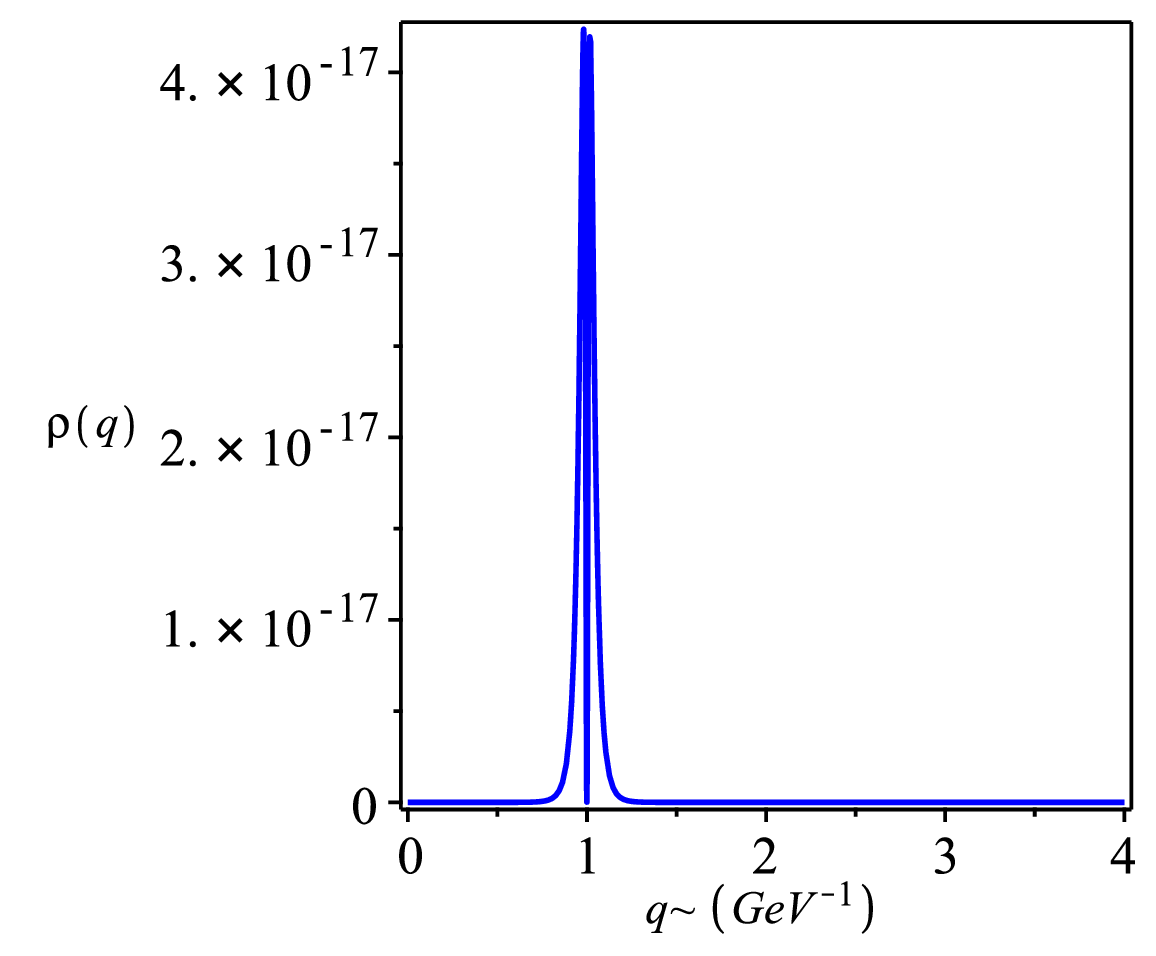}
\includegraphics[scale=0.30]{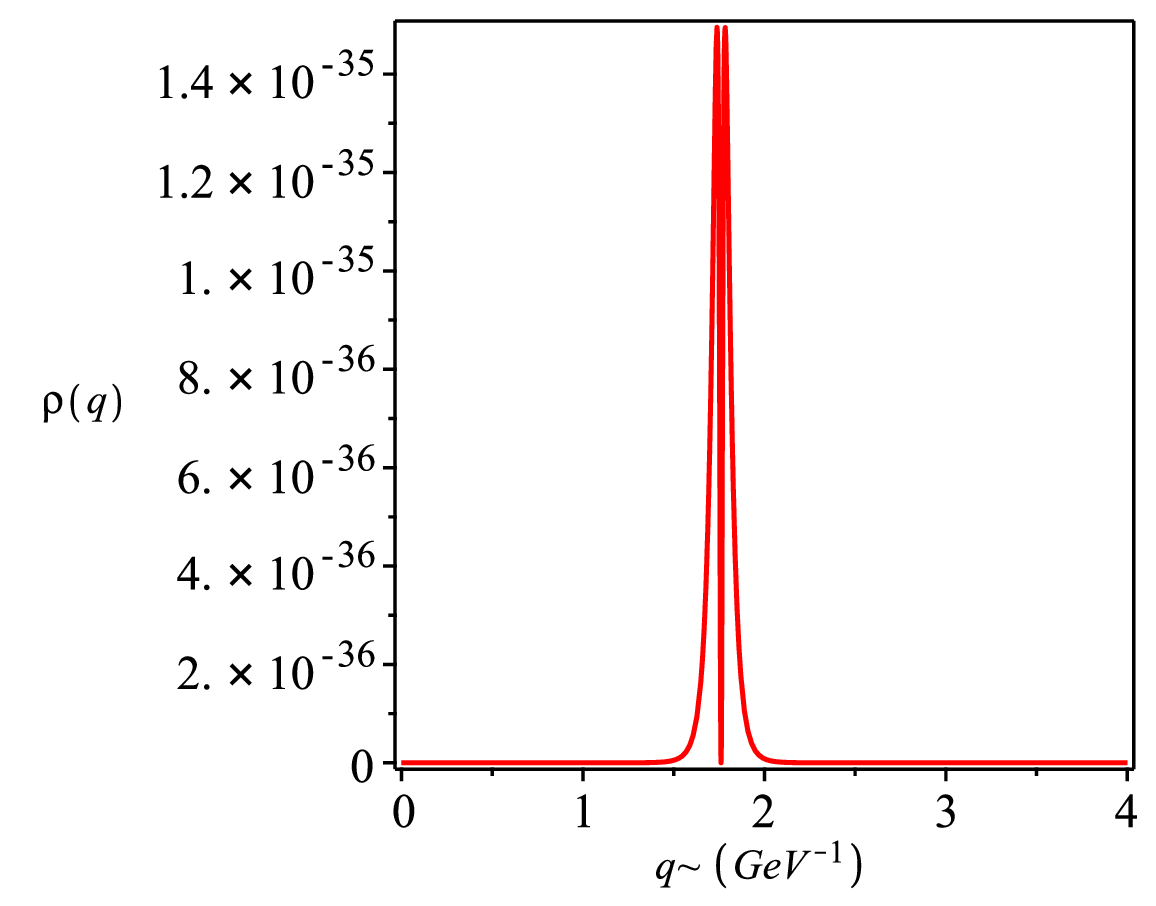}
\caption{The probability density obtained from the fractional Wigner function with fractional order $\alpha=0.155$ (black solid line) (meson $c\overline{c}$), $\alpha=0.155$ (blue solid line) (meson $b\overline{b}$), $\alpha=0.155$ ( red solid line)  and $\beta=0.5$ for the 1S state.}
\label{fig:wigner1}
\end{figure}

In this section, we analyze the behaviour of the fractional Wigner function for the fundamental energy level of $c\overline{c}$, $b\overline{b}$  and $b\overline{c}$. The following results are taking here~\cite{Mansour,RL_1}. For $c\overline{c}$:  $\hbar=1$, momentum $p=1$ $GeV$, reduced mass $m_{c\overline{c}}=0.73$ $GeV^{-1}$, confinement parameters $\lambda=0.3093$ $GeV^{2}$, $\sigma=0.1130$ $GeV^{3}$. For $b\overline{b}$: reduced mass $m_{b\overline{b}}=2.34$ $GeV$, confinement parameters $\lambda=0.2370$ $GeV^2$, $\sigma=0.1185$ $GeV^3$). For $b\overline{c}$: reduced mass $m_{b\overline{c}}=1.11$ $GeV$, $\lambda=0.1721$ $GeV^2$, $\sigma=0.0488$ $GeV^3$).  Figures~\ref{fig:wigner1} show the behavior solutions given in Eq.~\eqref{fracWigner}, for fractional order $\alpha$.

In Figure~\ref{fig:wigner1}, for fundamental level energy of system of $c\overline{c}$, $b\overline{b}$, and $b\overline{c}$ mesons, the probability density calculated from the fractional Wigner function, defined as $\int dp f_W = |\psi(q)|^2$, respectively, at fractional order, $\alpha$. In these systems, when compared to the experimental evidence, the value for the maximum relative distance where a quark-antiquark interact is approximately $4.077$ $GeV^{-1}$, for $c\overline{c}$; $2.58$ $GeV^{-1}$, for $b\overline{b}$  and $2.15$ $GeV^{-1}$, for $b\overline{c}$)~\cite{pdg,rluz,Mansour}.

The mass spectra of the heavy $c\overline{c}$ meson, heavy $b\overline{b}$   and $b\overline{c}$ were calculated. These mass spectra is given by the following equation~\cite{Mansour}
\begin{equation}
    M = m_{q} + m_{\overline{q}} + E'. \label{spectra}
\end{equation}
Then Eq.~\eqref{eq-e1} is used into Eq.~\eqref{spectra}, leading to

\begin{equation}
    M = m_{q} + m_{\overline{q}} + \frac{\lambda^2}{4\sigma}-\frac{\sigma A^2}{2m}\left [ n\alpha \sqrt{\frac{m}{4\sigma m^2}} + \sqrt{\frac{m}{\sigma A^2}\left ( \frac{1}{A^2}- \alpha^2 \right )}\right ] . \label{spectra1}
\end{equation}
Where $m_{q}$, $m_{\overline{q}}$ are the masses of the quark-antiquark and $m$ is the reduced mass. It is worth noting that we are  using the real energy eigenvalues only. In this case,  according to
condition given in Eq.~\eqref{eq-e2}, we take $\alpha=0.155$ and $\beta=0.5$ that appears to be more satisfactory to fit to the experimental data. In our calculations, we use the parameters, $m_{c}=1.4619$ $GeV$, $m_{b}=4.68$ $GeV$~\cite{Kumar,Nahool}. In Table~\ref{table:1}, the total error is calculated by averaging the relative errors with regard to the experimental data~\cite{pdg}.
Considering these values of $\alpha$ and $\beta$, the fundamental level energy and the mass spectra of the $c\overline{c}$, $b\overline{b}$ and $b\overline{c}$ mesons 1S-state are calculated, and presented in Table~\ref{table:1}. Therefore, the mass spectra of the $c\overline{c}$, $b\overline{b}$, $b\overline{c}$ mesons for 1S-state were compared with the 1S-state of the experimental data and other theoretical research carried out in the literature~\cite{AbuSh,Kumar,mabu1,Abdel,MAbu2016,Omugbe,rluz,pdg}.

In Ref~\cite{AbuSh}  predictions for both the 1S and 1S energies of the $c\overline{c}$, $b\overline{b}$ systems are presented by
a combination of methods, including extended Nikiforov-Uvarov method, generalized fractional derivative and the extended Cornell potential.
Those predictions for the 1S-state of the $c\overline{c}$ meson are $3.074$-GeV, and for the 1S-state $b\overline{b}$ meson, $9.465$-GeV with
fractional parameters $\alpha=0.84, \beta=1$. The predictions derived here for 1S $c\overline{c}$ meson are closest to the experimental result~\cite{pdg}. That is, for 1S-states for the $b\overline{b}$ and $b\overline{c}$ mesons there are discrepancies of about $0.003$-GeV, $0.004$-GeV with experiment.
While the findings of Ref.~\cite{AbuSh} diverge by $0.023$-GeV and $0.004$-GeV. Notice that Ref.~\cite{AbuSh} does not provide data for 1S
$b\overline{c}$ meson. Our results are compared with those of Ref.~\cite{Kumar}, who used the modified Cornell potential, asymptotic iteration
method together with N-dimensional radial Schrödinger equation. And those of Ref.~\cite{mabu1} that address a version extended Cornell
potential, the analytical exact iteration method in the N-dimensional space framework. We observed that for 1S-states of the $c\overline{c}$,
$b\overline{b}$ mesons, our predictions are more accurate than the models presented in Refs.~\cite{Kumar,mabu1} and exhibits a satisfactory
agreement when compared with experimental evidences~\cite{pdg}. The differences between the predictions for 1S-state $c\overline{c}$ system and
experiment~\cite{pdg} for this research and studied by~\cite{Kumar,mabu1} is $0.0004$-GeV, $0.019$-GeV and $0.0016$-GeV, respectively.

In the case of the 1S-state $b\overline{b}$ system, the margin of difference to the experiment results of this paper and the works~\cite{Kumar,mabu1} is
$0.0039$-GeV, $0.049$-GeV and $0.284$-GeV. For 1S-state $b\overline{c}$ system,  Ref.~\cite{Kumar} does not provide results,
meanwhile this paper and Ref.~\cite{mabu1} have a difference of $0.004$-GeV, $0.0004$-GeV. In Ref.~\cite{Abdel}, the spectra of heavy mesons and
its thermodynamic properties were obtained in the context of N-dimensional radial Schrödinger equation using extended Cornell potential.
Our predictions for the 1S-state $c\overline{c}$ meson improved the results presented in Ref.~\cite{Abdel} and are nearby the experimental. However,
measured value in Ref.~\cite{Abdel} for 1S-state $b\overline{b}$ meson is exact to experiment, while the differences between the present results and
Ref.~\cite{Abdel} with the experiment are $0.0004$-GeV, $0.002$-GeV for the 1S-state $b\overline{b}$.  Observe that Ref.~\cite{Abdel} does not
provide results for the 1S-state $b\overline{c}$ system. In Ref.~\cite{MAbu2016}, methods such as analytical exact iteration, N-dimensional radial
Schrödinger equation and extended Cornell potential are considered. The calculated value for the 1S-state $c\overline{c}$ system in this work are
consistently in good accord with the result of Ref.~\cite{MAbu2016}. However, for 1S-state $b\overline{b}$ and $b\overline{c}$ mesons the
predictions of  Ref.~\cite{MAbu2016} are improved with respect to experiment. Our measured mass spectra for 1S-states mesons
($c\overline{c}$, $b\overline{b}$ and $b\overline{c}$) differ with the experiment by $0.0004$-GeV, $0.003$-GeV, and $0.004$-GeV, while the
outcomes of work~\cite{MAbu2016} differ with the experiment by $0.001$-GeV, respectively.

Our results presented improvement for the 1S-state $b\overline{b}$ system than those the Ref.~\cite{Omugbe}, in which the authors obtained
mass spectra of heavy mesons under the spinless-Salpeter equation implemented with Cornell potential model and semi-classical WKB approximation
formulation. In addition, our predictions for 1S-states $c\overline{c}$ and $b\overline{b}$ system are more accurate compared to the Ref.~\cite{rluz}.
In a previous work~\cite{rluz} uses the formalism of symplectic quantum mechanics, generalized fractional derivatives and linear confining part of Cornell potential model.
These results suggests the validity of this works's model presented,  what reinforces the accounts formulated in section~\ref{FSE}.
Indeed, it is an improvement when compared to previous theoretical results~\cite{AbuSh,Kumar,mabu1,Abdel,Omugbe,rluz} and
exhibits a satisfactory agreement when compared with the experimental evidences~\cite{pdg}.

\begin{table}
\centering
\caption{Fundamental energy level and mass spectrum of mesons ($c\overline{c}$, $b\overline{b}$, $b\overline{c}$) in ($GeV$) with ($\alpha=0.155$ and $\beta=0.5$), ($\lambda=0.3093$ $GeV^2$, $\sigma=0.1130$ $GeV^3$, $m_{c\overline{c}}=0.73$ $GeV$), ($m_{b\overline{b}}=2.34$ $GeV$, $\lambda=0.2370$ $GeV^2$, $\sigma=0.1185$ $GeV^3$), ($m_{b\overline{c}}=1.11$ $GeV$, $\lambda=0.1721$ $GeV^2$, $\sigma=0.0488$ $GeV^3$)~\cite{Kumar,Nahool,pdg}.}
\label{table:1}
\begin{tabular}{c c c c c c c c c c c c}
 \hline
 Mesons & State & $E'$ & Present Work & Ref.~\cite{AbuSh} & Ref.~\cite{Kumar} & Ref.~\cite{mabu1} & Ref.~\cite{Abdel} & Ref.~\cite{MAbu2016} & Ref.~\cite{Omugbe} & Ref.~\cite{rluz} & Exp.~\cite{pdg} \\ [0.5ex]
 \hline\hline
 $c\overline{c}$ & $n=0$ & 0.1713 & 3.0966 & 3.074 & 3.078 & 3.0954 & 3.095 & 3.096 & 3.098 & 3.1003 & 3.097 \\

 $b\overline{b}$ & $n=0$ & 0.0964 & 9.4564 & 9.465 & 9.510 & 9.7447 & 9.460 & 9.460 & 9.681 & 9.4818 & 9.4603 \\

 $b\overline{c}$ & $n=0$ & 0.1311 & 6.2730 & $--$ & $--$ & 6.2774 & $--$ & 6.277 & $--$ & $--$ & 6.277 \\

 Total Relative Error &        &        &   $0.03\%$ &   $0.39\%$  &   $0.56\%$ & $1.02\%$ & $0.11\%$ &   $0.01\%$    & $11.08\%$ & $0.16\%$    &    $--$   \\ [1ex]
 \hline\hline
\end{tabular}
\end{table}

\begin{table}[h!]
\centering
\caption{The 1S-state mass spectra (in GeV) for $c\overline{c}$ system at different values of generalized fractional parameters $\alpha$ and $\beta$. The parameters ($\lambda$, $\sigma$, $m_{c\overline{c}}$) are the same as those used in Table~\ref{table:1}.}
\label{table:2}
 \begin{tabular}{c  c  c  c  c  c}
 \hline
 $c\overline{c}$ system & State & $E'$ & Present Work & Exp.~\cite{pdg} \\ [0.5ex]
 \hline\hline
 ($\alpha=0.15$), ($\beta=0.5$) & $n=0$ & 0.1713 & 3.0947 &  3.097\\
 ($\alpha=0.13$), ($\beta=0.5$) & $n=0$ & 0.1638 & 3.0872 & $--$ \\
 ($\alpha=0.2$), ($\beta=0.5$) & $n=0$ & 0.1896 & 3.1130 & $--$ \\
 ($\alpha=0.5$), ($\beta=0.5$) & $n=0$ & 0.2749 & 3.1983 & $--$ \\
 ($\alpha=0.15$), ($\beta=0.15$)& $n=0$ & 0.3635 & 3.2869 & $--$ \\
 ($\alpha=0.5$), ($\beta=0.15$) & $n=0$ & 0.5723 & 3.4957 & $--$ \\ [1ex] 
 \hline\hline
 \end{tabular}
\end{table}

Our results for 1S-state $c\overline{c}$ system are presented in Table~\ref{table:2} at different fractional orders $\alpha$, $\beta$ and
compared with experimental results from the Particle Data Group~\cite{pdg}. It is important to observe that for certain values of $\alpha$ and $\beta$ generalized
fractional parameters,  there is an increasing in the predicted mass spectra in comparison with experiment. The fractional parameters
choice becomes favorable for $\alpha=0.15$ and $\beta=0.5$. Whereas the generalized fractional parameters that giving a better fit to the experiment data
are those listed in Table~\ref{table:1}. 

\section{Final concluding remarks} \label{fr}

In the present paper, we have employed the generalized fractional Nikiforov-Uvarov method to find the quasi-amplitude solution for
heavy quark-antiquarks:charm ($c$)-anticharm ($\overline{c}$), bottom ($b$)-antibottom ($\overline{b}$), and $b-\overline{c}$ mesons in the formalism of the symplectic quantum mechanics with the Cornell potential expanded in the Coulonb-like term (a gluon contributions) up to the quadratic term in the distance of separation among the quarks in the meson.

The corresponding fractional Wigner function and the energy eigenvalues have been derived.
The effect of the fractional parameter $\alpha$  is addressed, considering the Wigner function in background of the formalism of symplectic quantum mechanics, an aspect not explored previously. The fractional parameters have relation with the existence of the Wigner function, which describes the meson fields. In addition, we note that the mass spectra of the heavy mesons are improved compared to other reports in the literature. The finite temperature effect and baryonic chemical potential~\cite{soleiman} are aspects to be addressed elsewhere.

\section*{Acknowledgments}

This work is partially supported by the Brazilian Government Agencies CNPq and CAPES for financial support.

\section*{Data Availability}
The authors declare that the data supporting the findings of this study are available within the paper and its references.

\end{document}